\documentclass[preprint2]{aastex}
\usepackage{graphicx}

\shorttitle{}
\shortauthors{Masciadri \& Raga}

\begin{document}


\title{Looking for outflows from brown dwarfs}

\author{E. Masciadri\altaffilmark{1}}
\affil{Max-Plank Institut f\"ur Astronomie,
K\"onigstuhl 17, D-69117 Heidelberg, Germany}
\email{e-mail: masciadr@mpia.de}
\author{A.C. Raga\altaffilmark{2}}
\affil{Ciencias Nucleares, UNAM, Mexico D.F., Mexico}




\begin{abstract}
First evidences of IR excess and disk mass accretion 
(strong H$\alpha$ emission) around 
brown dwarfs seem to indicate the existence of circumstellar disks around
these sub-stellar objects. Nothing is known at the present time about
outflows which potentially might be launched from brown dwarfs,
although jets are typically associated with the 
accretion in standard T~Tauri star disks.
In this paper we calculate the H$\alpha$ emission of internal 
working surfaces produced by a radiative jet in a neutral and in a 
photoionized 
environment as a function of the jet parameters
(the ejection velocity $v_j$, shock velocity $v_s$, mass loss rate ${\dot M}$
and radius $r_j$ of the jet) and we provide estimates of the H$\alpha$ luminosity
for the parameters of ``standard'' Herbig-Haro (HH) jets from T Tauri stars and
for the parameters expected for jets from BDs. Interestingly, we find that
while the mass loss rates associated with jets from BDs are found to be two
orders of magnitude lower than the mass loss rates associated with ``standard''
HH jets (from T~Tauri stars), their velocities are likely to be similar.
Based on our calculations, we discuss the conditions
in which jets from BDs can be detected and we conclude that the H$\alpha$ luminosities of internal 
working surfaces of jets from BDs in a photoionized environment
should have only one order of magnitude lower than the H$\alpha$ 
luminosities of
T~Tauri jets in a neutral environment.
\end{abstract}




\section{Introduction}
\label{intr}

Following the discovery (Nakajima et al. 1995) of the first brown dwarf (BD), 
several others have been detected in the past few years (Mart\'\i n et al. 1997; Kirkpatrick et al. 1999, 2001; Gizis et al. 2001; Burgasser et al. 2003; 
Gizis et al. 2003). Many of the properties of BDs are still unknown,
and even the definition of a brown dwarf remains a subject of discussion.
One of the fundamental open questions related to BDs
is their origin. Are they the result of a
molecular cloud collapse followed by the accretion 
of circumstellar disks formed around a central nucleus (Elmegreen 1999)
or are they stellar embryos 
that, due to dynamical interactions, are ejected from a multiple system before
having time to accrete sufficient
material from a circumstellar disk to reach the hydrogen burning limit (Reipurth \& Clarke 2001;
Bate, Bonnel \& Bromm 2003) ? 

It is even more difficult to define which 
are the observational evidences that could be used to discern between the two 
models. At the present time, there are no elements that can exclude one of these models, and it is quite important to collect observational evidence and/or theoretical arguments that can support or reject one or more of these models.
It is of course also possible that more than one of the proposed mechanisms
might be active in BD formation.

What about the existence of jets from brown dwarfs ? The existence of disks is a necessary condition for the existence of jets.
Evidence of accretion disks around BDs was recently obtained by IR excess
estimations (IR excesses being the typical signature of the presence of disks
around T~Tauri stars). Near-IR ($3.8$ $\mu$m) detections in IC348 and
Taurus (Liu et al. 2003), mid-IR ($7-14$ $\mu$m) detections in $\rho$ Oph and
Chamaeleon (Comer\'on et al. 1998; Comer\'on et al. 2000;
Natta \& Testi 2001; Testi et al. 2001; Natta et al. 2002) 
and sub-millimeter and millimeter detections
(Andr\'e \& Montmerle 1994; Carpenter 2002; Klein et al. 2003) 
started to identify
a rich sample of BDs with circumstellar disks. 
Besides this, signatures of mass outflows
are also obtained by the detection of strong H$\alpha$ 
emission principally in young
BDs (with ages of a few Myr) in the $\sigma$ Ori region
(Barrado y Navascu\'es et al. 2002;
Zapatero Osorio et al. 2002), Taurus (Mart\'\i n et al. 2001), and the 
TW Hydrae association (Gizis 2002).

It is more difficult to find a straight correlation between the presence of jets and the formation mechanism of BDs. In the scenario in which
BDs form from a direct collapse of a molecular cloud, they should have
an accretion disk similar to the one around a T~Tauri star, and therefore
the production of jets would be expected. In the scenario in which
BDs are stellar embryos ejected from a young, multiple system, the
BDs might still have low mass disks around them, which potentially
could also eject bipolar jet systems. Due to the fact that the
possible outflow ejection mechanisms associated with these different
BD formation models have not yet been studied in any detail, it is not
possible to say what a discovery of jets from BDs would imply about
the formation mechanism(s) of BDs. 

A paper that is particularly relevant in the context of jets from BDs
is the one of Quillen \& Trilling (1998), who showed that even
proto-Jupiters might be expected to produce jets. There is also
the work of Wolk \& Beck (1990), who extrapolated the outflow
luminosity vs. source mass correlation observed for jets from
T~Tauri (and more massive) stars to low stellar masses, to obtain
predictions of the luminosities expected for jets from BDs.


Which is the typical velocity of ejection of BD outflows ? Is the hydrodynamic
process producing shock emission in BD jets similar to the one of
HH jets from T~Tauri stars ? Which is the typical luminosity of their
emission ? Can we hope to detect BD jets ? What would be the best way to
search for them ? 

In the present paper, we discuss a simple, theoretical ``internal
working surface'' model, from which we obtain predictions of
the H$\alpha$ luminosity of an HH jet travelling within a neutral
medium or within a photoionized region (section 2). We use this 
model to obtain predictions of the H$\alpha$ luminosity of jets
from BDs (section 3). Finally, we use our results to propose observational
strategies for detecting jets ejected by BDs (section 4).

\section{An internal working surface}

\subsection{The basic geometry}

We model the knots along HH jets as ``internal working surfaces''
resulting from a time-variability of the ejection velocity 
(Raga et al. (1990), Raga \& Kofman (1992), Kofman \& Raga (1992)). 

\begin{figure}
\centering
\includegraphics[width=6cm]{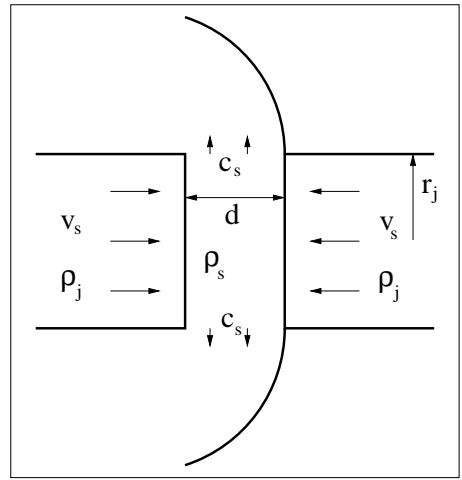}
\caption{Internal working surface (IWS) resulting from a variability of the
ejection velocity. The diagram shows the flow as seen from a reference
system moving with the IWS. In this reference system, the material in
the jet beam converges onto the IWS. If the up- and down-stream jet
beam segments have the same density $\rho_j$, the two IWS shocks have
the same shock velocity $v_s$. The shocked material within the working
surface (which has a density $\rho_s$ and sound speed $c_s$) escapes
sonically through the sides and interacts with the surrounding
environment, forming a bow shock. The radius of the jet is $r_j$,
and the separation between the two IWS shocks is $d$.
\label{fig1}} 
\end{figure}

In order to obtain analytic estimates of the emission expected from
an internal working surface (IWS), we consider the most simple possible
analytic model. The schematic diagram of Fig.\ref{fig1} shows the flow
structure of an IWS when seen from a frame of reference moving with
the velocity $v_{ws}$ of the working surface. In this frame of
reference, the material from the upstream and downstream continuous
jet segments converges onto the working surface, producing a
two-shock structure. 

The most simple situation is found for the case in which the
upstream and the downstream jet segments have the same density
$\rho_j$. In this case, the material in both beam segments converges
onto the working surface with the same velocity $v_s$ (see, e.~g.,
Raga \& Kofman 1992).

If the two shocks are strong and radiative, the shock jump conditions
can be written as~:
\begin{equation}
\rho_j{v_s}^2=\rho_s{c_s}^2\,,
\label{rj}
\end{equation}
where $\rho_j$ is the density of the continuous beam segments (see
Fig.\ref{fig1}), $v_s$ is the shock velocity, and $\rho_s$ and $c_s$ are, 
respectively, the density and isothermal sound speed ($\approx
10$~km~s$^{-1}$) 
at the end of the postshock cooling regions.

Following Falle \& Raga (1993), we can now estimate the separation
$d$ between the two IWS shocks (see Fig.\ref{fig1})
by assuming that the material escapes
sideways from the jet beam with a velocity equal to $c_s$, and then
write the inflowing/outflowing mass balance of the IWS~:
\begin{equation}
{\dot M}_{in}={\dot M}_{out}\,,
\label{mm}
\end{equation}
where ${\dot M}_{in}=2\pi {r_j}^2\rho_j v_s$ is the rate of mass
coming into the IWS through the two shocks, and ${\dot M}_{out}=
2\pi r_j d \rho_s c_s$, where $r_j$ is the jet radius (see Fig.\ref{fig1}).
From equation (\ref{mm}) we then obtain
that the separation $d$ between the two IWS shocks is given by:
\begin{equation}
d=\left({c_s\over v_s}\right)\,r_j\,.
\label{d}
\end{equation}


\subsection{Internal working surfaces in a neutral environment}

Let us now estimate the H$\alpha$ luminosity of an IWS of a jet
launched in a neutral region. 
For computing a simple estimate of the emission, we follow Raga \& Kofman
(1992), and assume that the emission comes mostly from the two shocks
within the jet beam, and that the emission from the
bow shock (which is pushed into the surrounding environment, see
Fig.\ref{fig1}) is negligible. 

The total H$\alpha$ luminosity of the IWS is given by~:
\begin{equation}
L_{H\alpha}^{(N)}=2\pi {r_j}^2 \sigma_{H\alpha}\,
\label{l1}
\end{equation}
where $\sigma_{H\alpha}$ is  the luminosity per unit area (of one of the shocks),
$r_j$ is the jet radius, and the contribution from the two IWS shocks
has been considered.

If one considers the H$\alpha$ emission
from the ``self consistent preionization'' plane-parallel shock
models of Hartigan, Hartmann \& Raymond (1987), one finds that
for the shock velocity range $v_s=30\to 100$~km~s$^{-1}$ the H$\alpha$
luminosity per unit area $\sigma_{H\alpha}$ of the shocks can be fitted
with the interpolation formula
\begin{equation}
\sigma_{H\alpha}={2\times 10^{-5}}\,{\rm erg\,cm\,s^{-1}}\,
\left({v_s\over {\rm 100\,km\,s^{-1}}}\right)^3\,n_j\,,
\label{sig}
\end{equation}
where $v_s$ is the shock velocity and $n_j$ ($\approx \rho_j/1.3m_H$
in our case, see Fig.\ref{fig1}) is the preshock number density.

We now write the average mass loss rate from the time-dependent
source as
\begin{equation}
{\dot M}=\pi{r_j}^2\rho_j v_j\,,
\label{m}
\end{equation}
where $v_j$ is an appropriately defined average value of the ejection
velocity. Combining equations (\ref{l1}-\ref{sig}-\ref{m}) we finally obtain
the H$\alpha$ luminosity
$$L_{H\alpha}^{(N)}={3.0\times 10^{-3}{\rm L_\odot}}
\left({{\dot M}\over {\rm 10^{-7}M_\odot yr^{-1}}}\right)\times
$$
\begin{equation}
\left({{\rm 100\,km\,s^{-1}}\over v_j}\right)
\left({v_s\over {\rm 100\,km\,s^{-1}}}\right)^3\,,
\label{ln}
\end{equation}
normalized to typical parameters for a ``normal'' HH flow from a
T~Tauri star.

\subsection{Internal working surfaces in a photoionized environment}
\label{}

Let us now estimate the H$\alpha$ luminosity of an IWS of a jet
immersed in a photoionized region. As has been discussed by Raga et al.
(2000) and by L\'opez-Mart\'\i n et al. (2001), in many cases HH jets
can be completely photoionized by the ionizing radiation field from the
H~II region source. We will therefore assume that we have an IWS
which is fully photoionized by the external, ionizing field.

We will then compute an estimate of
the H$\alpha$ emission coming from the volume
of gas in between the two IWS beam shocks (see Fig.\ref{fig1}) as~:
\begin{equation}
L_{H\alpha}^{(P)}=(\pi r_j^2 d) {n_s}^2\alpha_{H\alpha}(T_s)\,h\nu_{H\alpha}\,,
\label{l2}
\end{equation}
where the first factor in parentheses on the right is the volume of
emitting material (see Fig.\ref{fig1}), $n_s=\rho_s/(1.3\,m_H)$ is the
post-cooling region number density, $\alpha_{H\alpha}(T_s)$ is
the effective H$\alpha$ recombination coefficient (calculated
at the $T_s\approx 10^4$~K temperature of the post-cooling region
photoionized material) and $h\nu_{H\alpha}$ is the energy of the
H$\alpha$ transition. Using equations (\ref{rj}), (\ref{d}) and
(\ref{m}) and setting $c_s\approx 10$~km~s$^{-1}$ we then obtain
$$L_{H\alpha}^{(P)}={1.68\,{\rm L_\odot}}
\left({{\dot M}\over {\rm 10^{-7}M_\odot yr^{-1}}}\right)^2\times
$$
\begin{equation}
\left({{\rm 100\,km\,s^{-1}}\over v_j}\right)^2
\left({v_s\over {\rm 100\,km\,s^{-1}}}\right)^3
\left({{\rm 100\,AU}\over r_j}\right)\,,
\label{lp}
\end{equation}
normalized to typical parameters for a ``normal'' HH flow from a
T~Tauri star.

Actually, to this H$\alpha$ luminosity we should add the contribution
from the cooling regions right behind the two IWS shocks. However,
from a comparison of equations (\ref{lp}) and (\ref{ln}) we can see
that this contribution has to be negligible, at least for the parameters
of a ``normal'' HH jet.

Another interesting feature of fully photoionized jets is that
the emission from the jet beam itself can be quite substantial.
It is straightforward to show that for a constant density
jet the H$\alpha$ luminosity from the whole of the jet beam is
of the order of $L_{H\alpha}^{(P)}$ (the luminosity
of the working surface, see equation \ref{lp}) multiplied by
a factor $(L/r_j)(c_s/v_s)^3$ (where $L$ is the total length of the
jet beam and $c_s\approx 10$~km~s$^{-1}$ is the sound speed of
the photoionized gas). For example, for $L/r_j\sim 100$ and
$v_s\sim 100$~km~s$^{-1}$, we would have a jet beam H$\alpha$ luminosity
of 10~\%\ of the working surface luminosity. Though being quite
diffuse, this emission from the jet beam might also be detectable.

\section{Optical emission of outflows from brown dwarfs}

In this section we use the results obtained in 
sections 2.2 and 2.3 to try
to compare the emission properties of typical HH jets with the properties
of jets ejected by brown dwarfs.

All of the jet (or wind) ejection mechanisms have in common the fact
that the ejection velocity $v_j$
is of the order of the escape velocity $v_{es}$ ($v_j$ $\sim$ $v_{es}$) 
from the surface of
the outflow source. For an outflow from a T~Tauri star we then have that:
\begin{equation}
v_{es,TT} = \left(\frac{2GM_{TT}}{r_{TT}}\right)^{0.5}\,,
\end{equation}
where $M_{TT}$ is the mass and $r_{TT}$ the radius of a T~Tauri star
and $G$ is the gravitational constant. For
a BD, we will have an outflow with a velocity of the order of the escape
velocity
\begin{equation}
v_{es,BD}=v_{es,TT}\cdot\left({\frac{M_{BD}\cdot r_{TT}}
{M_{TT}\cdot r_{BD}}}\right)^{0.5}\,,
\label{mass_rad}
\end{equation}
where $M_{BD}$ is the mass and $r_{BD}$ the radius of a BD.

Which are the typical values for $M_{BD}$ and $r_{BD}$ ? Under the
assumption that the BDs which are most likely to produce observable
jets are the youngest and most massive ones we can reasonably estimate a
range of $M_{BD}$ $\sim$ $0.03-0.075$ M$_\odot$ for the mass.
It is more difficult to identify a typical value for $r_{BD}$.
From cool, sub-stellar object atmosphere models (Baraffe et al. 1998;
Baraffe et al. 2003; Burrows et al. 1997; Burrows et al. 2001)
one sees that for old objects (older than $\sim$ $1$ Gyr) the radius of
brown dwarfs $r_{BD}$ converges to a typical value of $\sim$ $0.1$ $R_\odot$.
For younger BDs (of a few Myr), $r_{BD}$ varies within the $0.1-0.4$ $R_\odot$
range.

If we then consider typical values $r_{TT}$ $\sim$ $4$ $R_\odot$,
$M_{TT}$ = $1$M$_\odot$ for a T~Tauri star, and $M_{BD}$ = $0.06$M$_\odot$,
$r_{BD}\sim 0.3$ $R_\odot$ for a BD, from equation (\ref{mass_rad})
we obtain that $v_{es,BD}$ $\sim$ $v_{es,TT}$. 
Therefore, using typical masses
and radii, we find that the escape velocities from the
surfaces of BDs have values that are similar to the ones
of T~Tauri stars. We conclude that the
velocities $v_{j,TT}$ and $v_{j,BD}$ of ejection from T Tauri stars 
and BDs are also comparable.

Another important parameter is of course the shock velocity $v_s$
of the shocks associated with the IWS (see, equations \ref{ln} and
\ref{lp}). This shock velocity is not only important for determining
the H$\alpha$ luminosity of the knots, but it also determines the
excitation/ionization of the emitted spectrum. In models of outflows
from time-dependent sources the shock velocity $v_s$ has a maximum
possible value $v_s$ $\sim \Delta v_j/2$ (where $\Delta v_j$ is the
full amplitude of the ejection velocity variability).


As no clear model for the production of ejection variabilities has
yet been studied in sufficient detail, it is not possible to find
a theoretical relation between the ejection velocity $v_j$ and the
amplitude $\Delta v_j$ of the ejection velocity time-variability.
However, from observations of HH jets (see, e.~g., the compilation
of Raga, B\"ohm \& Cant\'o 1996) it is clear that the shock velocities
of the shocks associated with these objects range from $v_{s,min}$
$\sim 20$~km~s$^{-1}$
(this lower limit probably being due to the fact that lower velocity
shocks produce very faint line emission and are probably not
detected) up to $v_{s,max}\sim 100$~km~s$^{-1}$. Therefore, the amplitude
of the velocity variability ranges from low values up to
$\Delta v_{j,max}\sim 2\,v_{s,max}\sim 200$~km~s$^{-1}$. In other words,
the amplitude of the velocity variability reaches values similar to the
mean velocity $v_j$ of the outflow. 
It appears thus to be reasonable to assume that this result might also be applicable
for jets ejected by BDs rather than by T~Tauri stars.


Therefore, jets from brown dwarfs have probably
knots with emission spectra of similar excitation/ionization
to the ones of jets from T~Tauri jets. Also, the H$\alpha$ luminosities
of the knots will be given by equations (\ref{ln}) and (\ref{lp}) with
values of $v_j$ and $v_s$ similar to the ones given for T~Tauri jets.

However, the H$\alpha$ luminosity is strongly affected by the fact
that the accretion rates onto young brown dwarfs appears to be roughly
two orders of magnitude lower than the accretion rates onto T~Tauri
stars (Muzerolle et al. 2003; Barrado y Navascu\'es et al. 2004).
One can roughly consider a range of ${\dot M}_{TT}$ $\sim$ $10^{-8}-10^{-7}$
M$_\odot$ yr$^{-1}$ for T-Tauri stars and a range of
${\dot M}_{BD}$ $\sim$ $10^{-10}-10^{-9}$
M$_\odot$ yr$^{-1}$ for brown dwarfs. As the mass loss rate in an
outflow appears to be proportional
to the mass accretion rate (Cabrit et al. 1990), this result directly implies
that the mass loss rate of a brown dwarf jet is at least two orders of
magnitude lower than the mass loss rate of a typical T~Tauri jet.

This lower mass loss rate implies that the H$\alpha$ luminosity
of a jet from a BD embedded in a neutral region will be
two orders of magnitude lower than the H$\alpha$ luminosity of
a typical jet from a T~Tauri star (see equation \ref{ln}).
Taking values of $v_j$=$100$ km s$^{-1}$, $v_s$=$100$ $km$ $s^{-1}$
 and ${\dot M}$=$10^{-9}$ M$_\odot$ yr$^{-1}$ 
the knots along a brown dwarf jet in a neutral region should have H$\alpha$
luminosities $L_{H\alpha}^{(N)}(BD)\sim {3\times 10^{-5}}$~L$_\odot$ (see
equation \ref{ln}).

It is not possible to make such a firm statement for the case
of a BD jet embedded in a photoionized region. The first
thing that we see from equation (\ref{lp}) is that a drop of two
orders of magnitude in the mass loss rate results in a drop of
four orders of magnitude in the H$\alpha$ luminosity (when going
from a T~Tauri to a BD jet). If we assume that the
radius of a BD jet has a value similar to the one
of a T~Tauri jet, we would conclude that the knots along a
photoionized BD jet should have an H$\alpha$ luminosity
$L_{H\alpha}^{(P)}(BD)\sim {1.7\times 10^{-4}}$~L$_\odot$ (see
equation (\ref{lp})). 

\section{Conclusions}

We have obtained simple, analytic estimates of the H$\alpha$ luminosities
of internal working surfaces of jets in a neutral environment 
(see section 2.2) and of fully photoionized jets (see section 2.3). 

We have argued that the dynamical characteristics (i.~e., jet velocity
and amplitude of the ejection velocity variability) are likely to be similar
for jets from T~Tauri stars and jets from BDs (see section 3).
Therefore, the lower observed mass accretion rates (which in its turn
determine the mass outflow rates) directly imply that ``neutral''
brown dwarf jets should have H$\alpha$ luminosities
two orders of magnitude lower than the luminosities
of ``normal'' HH jets from T~Tauri stars. 
This result is in agreement with the H$\alpha$ luminosities
calculated for BD jets by Wolk \& Beck (1990), who extended
to low source masses the $L_{H\alpha}$ vs. $M_{source}$ correlation
found for outflows ejected from T~Tauri (and more massive) stars.

We find that the observed correlation can be interpreted in terms of
our internal working surface model as a sequence of outflows with
the same jet velocity $v_j$ and variability amplitude $\Delta v_j$,
but with different mass loss rates ${\dot M}_j$.


We find that the knots along photoionized BD
jets have H$\alpha$ luminosities $L_{H\alpha}^{(P)}(BD)\sim
{10^{-4}}$~L$_\odot$ (see section 4). Even though
such luminosities are 4 orders of magnitude lower than the ones
of photoionized T~Tauri jets (see equation \ref{lp}), they are
only lower by an order of magnitude with respect to 
the luminosities of ``neutral'' T~Tauri jets
(see equation \ref{ln}).

This result indicates that favourable conditions 
to carry out searches
for BD jets can be found in regions with young, low mass stars embedded
within an H~II region, such as M~42 (Bally \& Reipurth 2001),
the Pelican Nebula (Bally \& Reipurth 2003) or the IC 1396N
H~II region (Reipurth et al. 2003). 
However, as is clear from the published images, in such
regions it is difficult to detect jets against the bright and inhomogeneous
background of the H~II region (which many times has ridges or bright
edges which might or might not be jets).
This problem could be
overcome by searching for photoionized BD jets with a large
field Fabry-P\'erot interferometer, in which even faint jets should
clearly stand out from the H~II region emission at high radial velocities.

As a final comment, let us point out that in order to fully
photoionize an HH jet, it is necessary to have a relatively strong
and nearby source of photoionizing radiation. For producing an
H$\alpha$ luminosity of $\sim 1\,L_\odot$ (as predicted for a jet
from a T~Tauri star, see equation 9), it is necessary to have
$\sim 5\times 10^{45}$~s$^{-1}$ ionizing photons. In order for a working
surface to intercept this number of photons, it is necessary to
have something like an O5 star (producing $S_*\approx {5\times
10^{49}}$~s$^{-1}$
ionizing photons) within a distance of $\sim 50\,r_j$ or
an O9 star (with $S_*\approx {2\times 10^{48}}$~s$^{-1}$)
within a distance of $\sim 20\,r_j$. The same stars could
fully photoionize the much more puny BD jets at distances
a factor of $\sim 100$ times larger than the ones quoted above.

\acknowledgments

E. Masciadri gratefully acknowledges the hospitality of the Instituto de Ciencias Nucleares, UNAM, Mexico D.F. during this work. The work of A.C. Raga was supported by CONACyT grants 36572-E and 41320 and the DGAPA (UNAM) grant IN 112602. We acknowledge the referees (Bo Reipurth and Noam Soker) for valuable commentaries.

\end{document}